%% file: HI_ring.tex
\documentclass[fleqn,usenatbib]{mnras}

\usepackage{newtxtext,newtxmath}

\usepackage[T1]{fontenc}
\usepackage{ae,aecompl}
\usepackage[utf8]{inputenc}
\usepackage{graphicx}	
\usepackage{rotating}
\usepackage{amsmath}	
\usepackage{amssymb}	
\newcommand{\aips}{\textsc{AIPS}}
\newcommand{\etal}{et al.\ }

\newcommand{\hi}{H{\sc i}}

\newcommand{\ATLAS}{ATLAS$^{\mathrm{3D}}$}

\title[Large \hi{} ring]{Discovery of a large \hi{} ring around the quiescent galaxy AGC 203001}

\author[Bait \etal]
{Omkar Bait,$^{1}$\thanks{E-mail: omkar@ncra.tifr.res.in (OB)}, Sushma Kurapati$^{1}$, Pierre-Alain Duc$^{2}$,
Jean-Charles Cuillandre$^{3}$,\newauthor
Yogesh Wadadekar$^{1}$, Peter Kamphuis$^{4}$ and Sudhanshu Barway$^{5}$
    \\
$^{1}$National Centre for Radio Astrophysics, Tata Institute of Fundamental Research, Post Bag 3, Ganeshkhind, Pune
411007, India \\
$^{2}$Universit{\`e} de Strasbourg, CNRS, Observatoire astronomique de Strasbourg, UMR 7550, F-67000 Strasbourg, France\\
$^{3}$AIM, CEA, CNRS, Universit{\`e} Paris-Saclay, Universit{\`e} Paris Diderot, Sorbonne Paris Cit{\`e}, Observatoire de Paris, PSL University,\\ 91191 Gif-sur-Yvette Cedex, France\\
$^{4}$Astronomisches Institut Ruhr-Universit\"at Bochum (AIRUB), Universit\"atsstrasse 150, D-44780 Bochum, Germany\\
$^{5}$Indian Institute of Astrophysics (IIA), II Block, Koramangala, Bengaluru 560 034, India}

\date{Accepted XXX. Received YYY; in original form ZZZ}

\pubyear{2019}

\begin{document}

\pagerange{\pageref{firstpage}--\pageref{lastpage}}
\maketitle



\begin{abstract}
Here we report the discovery with the Giant Metrewave Radio Telescope of an extremely large ($\sim$115 kpc in diameter)  \hi{} ring off-centered from a massive quenched galaxy, AGC 203001. This ring does not have any bright extended optical counterpart, unlike several other known ring galaxies. Our deep $g$, $r$, and $i$ optical imaging of the \hi{} ring, using the MegaCam instrument on the Canada-France-Hawaii Telescope, however, shows several regions with faint optical emission at a surface brightness level of $\sim$28 mag/arcsec$^2$.  Such an extended \hi{} structure is very rare with only one other case known so far -- the Leo ring. Conventionally, off-centered rings have been explained by a collision with an ``intruder" galaxy leading to expanding density waves of gas and stars in the form of a ring. However, in such a scenario the impact also leads to large amounts of star formation in the ring which is not observed in the ring presented in this paper. We discuss possible scenarios for the formation of such \hi{} dominated rings.

\end{abstract}



\section{Introduction} \label{intro}
Ring galaxies have been studied in the optical for a long time, starting with the now famous ``Cartwheel" ring \citep{Zwicky41}. This ring harbours a wide range of features like a bright blue ring, spokes, an inner secondary ring and a nucleus. The first breakthrough in understanding their origin came from \citet{Lynds76} and \citet{Theys77} who showed that such rings could form due to a density wave from a head-on high speed collision with an intruder galaxy. Such rings, which are classified as P-type rings (also termed as collisional rings), typically have an off-centre nucleus and a knotty structure \citep{Few86}. They differ from another class of ring galaxies, termed as O-type rings, which have a central nucleus and a smooth profile (\cite{Few86}) and are believed to have formed due to Lindblad resonances \citep{deVaucouleurs59, Binney08}. The strong impact in collisional rings leads to a burst of star formation (SF) in the whole system \citep{Appleton87, Struck-Marcell87}. It was shown in a recent simulation of a Cartwheel--like galaxy, that the location of active SF follows a particular sequence, wherein SF is first enhanced in the ring shortly after the interaction, then in the spokes and finally in the nucleus \citep{Renaud18}. Collisional rings are also known to form knots due to their self-gravity, a phenomenon known as ``bead instability" \citep{Dyson1893}, as first pointed out by \citet{Theys77} and also seen in a more detailed simulation by \citet{Hernquist93}. This leads to the fragmentation of the ring in about  100 Myr \citep{Theys77} and it fades almost completely in about 0.5 Gyr \citep{Mapelli08} thus making the ring fairly short-lived. Interestingly, such collisional rings are also produced in large scale cosmological simulations whose SF properties are found to be consistent with observations \citep{Snyder15, Elagali18}.

 It is well known that early-type galaxies (ETGs) can host \hi{} \citep[e.g.,][]{vanDriel91, Oosterloo02, Morganti06, Oosterloo10, Serra12}, even more so in non-cluster environments, in a variety of morphologies ranging from settled, unsettled discs/rings to disturbed morphologies \citep{Serra12}. In cases where large amounts of \hi{} is hosted by ETGs, it is known to be distributed in the form of large ring/disc morphology \citep[e.g.,][]{Morganti03a, Serra12} with signs of recent star-formation in it \citep{Yildiz17}. Among them, there is one case of an optically devoid large \hi{} ring -- the `Leo ring' \citep{Schneider83, Schneider85, Schneider89}.  Here, the \hi{} is distributed in the form of a spectacularly large and clumpy ring of about $200$~kpc diameter around the ETG pair NGC 3384 and NGC 3379 (M105), and has no obvious ultra-violet (UV)/optical counterpart. It was thus long thought to have a primordial origin \citep{Schneider83, Schneider85,Schneider89}. However, recent observations have found UV \citep{Thilker09}, optical \citep{Michel-Dansac10}, weak dust emission \citep{Bot09} and also metal-line absorption in the ring \citep{Rosenberg14}.  \citet{Stierwalt09} in their  Arecibo Legacy Fast ALFA (ALFALFA) observations of the Leo I group also identified about six candidate optical counterparts to the ring. However, only two of them, AGC 202027 and AGC 201970, had optical redshifts (from Sloan Digital Sky Survey (SDSS) and \citet{Karachentsev04} respectively) close of the \hi{} emission, and are thus associated with the ring. One of them, AGC  205505, has a slightly higher redshift than the \hi{} emission from the ring, and is thus likely to be a background galaxy, although still a part of the M96 group \citep{Stierwalt09}. The rest of the three galaxies  AGC 201972 (also called KK94 in \citet{Karachentseva98}) and AGC 202026, and AGC 201975 do not have redshift information to confirm the association with the ring. The metallicity of the Leo ring was found to be pre-enriched \citep{Michel-Dansac10, Rosenberg14} suggesting a galactic origin.  Recently, a very diffuse stellar object (BST1047+1156) was found in the region connecting the Leo ring and M96 which could either be a tidal dwarf galaxy or a pre-existing extremely diffuse low surface brightness (LSB) galaxy \citep{Mihos18}. Targeted simulations showed that the Leo ring could also have a collisional origin \citep{Michel-Dansac10}. An alternative scenario for the formation of such massive, optically devoid \hi{} clouds is the stripping of a gas-rich LSB galaxy with an extended \hi{} disc by the tidal field of galaxy groups \citep{Bekki05}.

 Here, we report the discovery of a large ($\sim$ 115 kpc in diameter) \hi{} ring off-centered around a massive quenched galaxy, AGC 203001 using the Giant Metrewave Radio Telescope (GMRT). We also present deep optical $g$, $r$, and $i$ band imaging of the \hi{} ring using the Canada France Hawaii Telescope (CFHT). In the context of this discovery, we discuss possible formation mechanisms for large \hi{} rings.

Throughout this paper, we use the standard concordance
cosmology from WMAP9 \citep{Hinshaw13} with $\Omega_M = 0.286$, $\Omega_\Lambda = 0.714$ and $h_{100} =
0.69$.

\section{Target selection}
Our target, AGC 203001 (RA: 10h22m03.315s, DEC: +13d46m10.20s) belongs to a sample of quenched galaxies selected in the following manner. We first select galaxies with specific star formation rate (star formation rate/stellar mass) below 10$^{-11}$ yr$^{-1}$ from the Galaxy Evolution Explorer (GALEX)-SDSS-Wide-field Infrared Survey Explorer (WISE) Legacy Catalog \citep[GSWLC;][]{Salim16}. Such galaxies are designated as quenched galaxies \citep{Salim2014}. Subsequently, we put a redshift cut of $z < 0.025$ on this sample to ensure sufficient spatial resolution with the GMRT and then we cross-match this sample with the optical counterparts identified from the single dish ALFALFA \citep[][]{Haynes18} \hi\ survey.
This resulted in a sample of 24 galaxies, of which 12 galaxies had early-type morphology and showed no signs of recent star-formation or interactions. From these, AGC203001 was picked at random as a pilot object for observing the full sample with the GMRT. AGC 203001 is at a redshift of 0.01867 (v$_{\mathrm{helio}}= 5597$\ km/s), has a stellar mass (M$_*$) of  $1.5 \times 10^{10} \ \text{M}_\odot$ \citep{Salim16} and a S0 morphology \citep{Nair10}. It has a SF rate (SFR) of 0.011 $\text{M}_\odot$/yr which is very low for its stellar mass, and hence the galaxy is considered quenched. Yet, it has copious amounts of \hi{} ($\log(\mathrm{M}/\mathrm{M}_\odot) = 9.53 \pm 0.05$) as measured by the single dish ALFALFA survey.

\section{Observations and data analysis}
\subsection{Deep \hi{} imaging of AGC 203001}
The GMRT 21 cm line observations of AGC 203001 were carried out on 23rd and 24th December 2017 for a total on-source time of $\sim$10 hours with the GMRT software backend. The total bandwidth was configured at 16.66 MHz divided into 512 channels, thus corresponding to a velocity resolution of $\sim$7 km/s, and centered at the heliocentric redshift of AGC 203001. The data analysis was carried out using standard tasks in the Astronomical Imaging Processing System \citep[{\sc aips} ;][]{Wells85} package. The initial flagging and calibration was done manually. Then a continuum image was made from the line free channels which was used for self-calibration. After applying the self-calibration solution to the line visibilites, the continuum emission was subtracted using the task {\sc uvsub}. The cube was then imaged using the task {\sc IMAGR}, and residual continuum was subtracted out using the task {\sc imlin}. We reach a noise level of $\sim$ 0.67 mJy/beam for a velocity resolution of 14 km/s and with a beam size of 38.02\arcsec{} $\times$ 35.24\arcsec{}. The \hi{} integrated intensity image and velocity field were made using the task {\sc momnt}.

\subsection{Deep optical imaging of the \hi{} ring}
On the publicly available images from optical imaging surveys such as  DeCALS or PanSTARRS, the HI ring does not show any obvious optical counterpart.
To check whether faint emission could nevertheless be associated, we have obtained  deep multi-band optical images of the field around AGC 203001 with the MegaCam camera installed on the Canada-France-Hawaii Telescope. The observation and data-reduction procedure which is optimised for the detection of low-surface brightness structures was the same as that used for the MATLAS/Atlas3D   survey \citep{Duc15}. Final images were obtained with the Elixir-LSB software combining 7 individual images acquired with relative offsets up to 20 arcmin, from which a master sky had been subtracted. Total exposure times were 34 min in the g' and r' bands, and 25 min in the i* band. Weather conditions were photometric.

\section{The \hi{} ring around AGC 203001}

\begin{figure*}
\begin{center}
\includegraphics[scale=0.35]{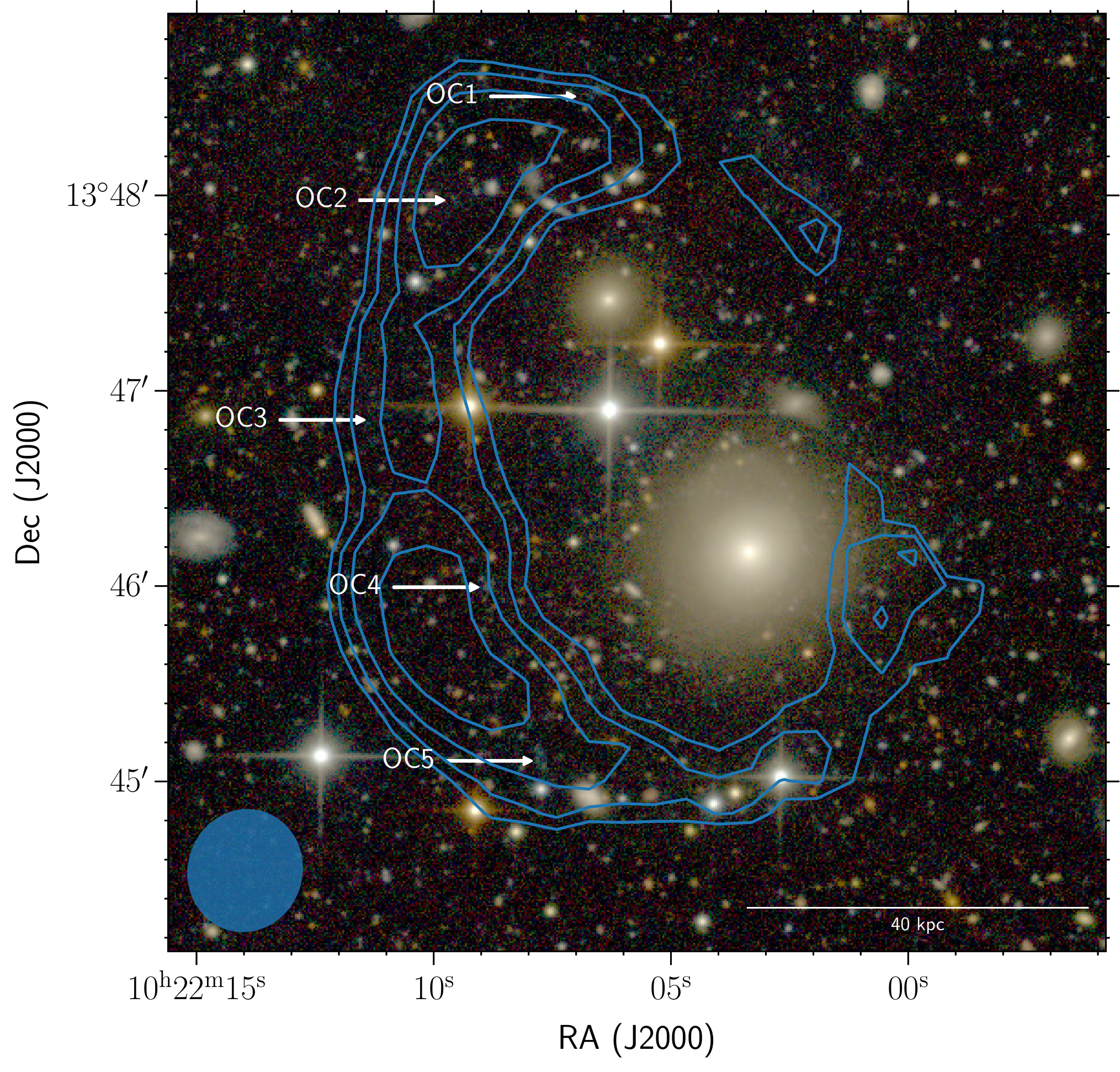} \includegraphics[scale=0.35]{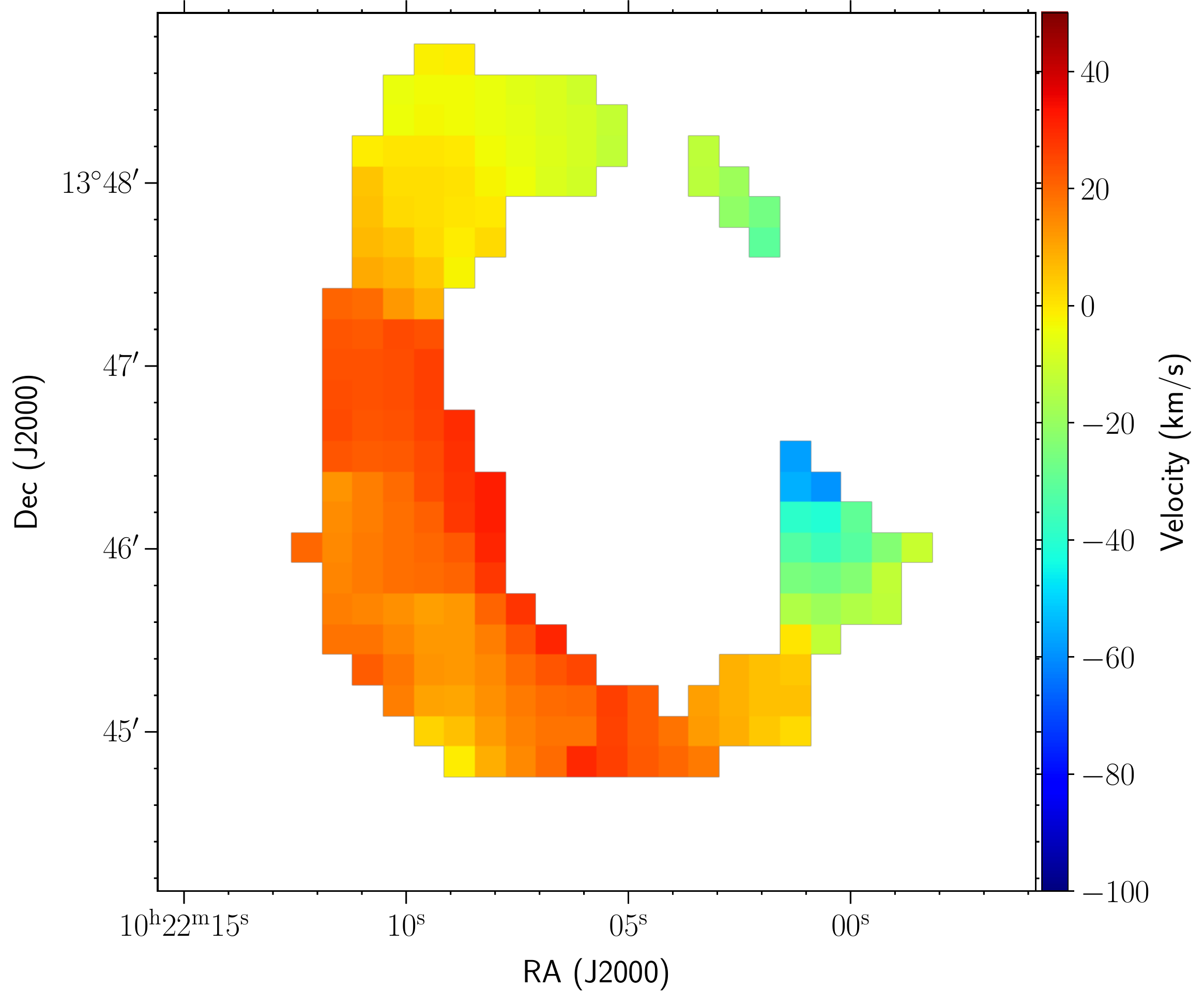} 
\caption{Left Panel: The \hi{} ring around AGC 203001 in blue contours overlaid on the CFHT optical $g$, $r$, $i$-band color composite image. The lowest contour corresponds to a \hi{} column density of 2.8 $\times 10^{19}$cm$^{-2}$ and subsequent contours rise in multiples of $\sqrt{2}$. The \hi{} ring has a projected diameter of $\sim115$ kpc. The arrows show the positions of the identified optical counterparts. Right Panel: The \hi{} velocity map of the ring.} 
\label{HI_overlaid}
\end{center}
\end{figure*}

 \begin{table}
  \begin{center}
    \caption{General properties of the \hi{} ring. }
    \scalebox{0.8}{
    \begin{tabular}{cc}
     
      \hline
      \hline
      Total detected \hi{} flux &  1.13 $\pm$ 0.13 Jy~km/s \\
      Systematic velocity & 5602.54~km/s \\
      Velocity width & 28.52~km/s (W50), 99.80~km/s (W20)\\
      Ellipticity & 0.67\\
      Position angle of the major-axis & 20$^{\circ}$\\
      Peak column density & 1.1$\times 10^{20}$~cm$^{-2}$\\
      \hline
    \end{tabular}
    } 
    \label{tab:HI_properties}
  \end{center}
\end{table}

\begin{figure*}
\begin{center}
\includegraphics[scale=1]{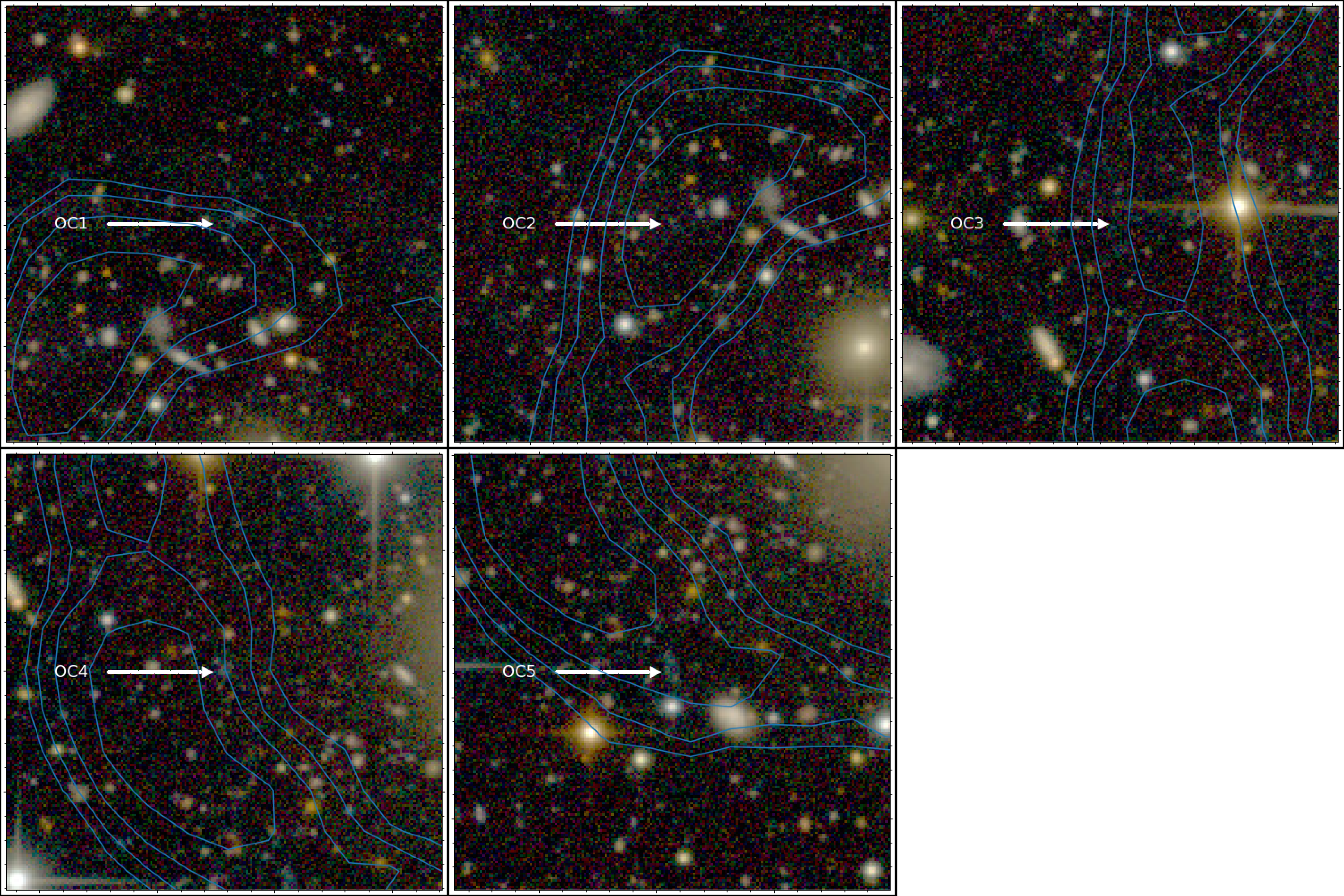}
\caption{A 1.8\ $\times$\ 1.8 arcmin$^2$ zoom-in near the various optical counterparts of the ring. The \hi{} column density contours are repeated from Fig. \ref{HI_overlaid}.} 
\label{OC_montage}
\end{center}
\end{figure*}

\begin{table*}
  \begin{center}
    \caption{ Positions and magnitudes of optical counterparts to the ring} 
    \begin{tabular}{cccccc}
     
      \hline
      ObjiD & RA & Dec & $g$ & $r$ & $i$\\ 

       & J2000 (deg) & J2000 (deg)  &  (mag) & (mag) & (mag)\\
       \hline
       OC1 & 155.52844 & 13.80845  & 23.65 $\pm$\ 0.18 & 23.04 $\pm$\ 0.15 & -- \\
      \hline
      OC2 & 155.53990 & 13.79961  & 24.81 $\pm$\ 0.40 & -- & -- \\
      \hline
      OC3 & 155.54692 & 13.78085  & 25.65 $\pm$\ 0.91 & -- & -- \\
      \hline
      OC4 & 155.53694 & 13.76658  & 23.33 $\pm$\ 0.12 & 23.26 $\pm$\ 0.17 & 22.98 $\pm$\ 0.28 \\
      \hline
      OC5 & 155.53220 & 13.75176  & 23.51 $\pm$\ 0.14 & 23.21 $\pm$\ 0.13 & -- \\
      \hline
      \hline
    \end{tabular}
    \label{tab:fluxes}
  \end{center}
\end{table*}

We find that the \hi{} in this galaxy is distributed in the form of a large ring around it, with a diameter of $\sim$115 kpc, as shown in blue contours in the left panel of Fig. \ref{HI_overlaid}, with the optical $g$, $r$, and $i$-band colour composite image from CFHT in the background. Notice that the ring is asymmetrically distributed around the host galaxy. The right panel in Fig.~\ref{HI_overlaid} shows the \hi{} velocity map of the ring centered on AGC 203001, where we see that the ring shows some gradient in velocity. Table. \ref{tab:HI_properties} describes some of the general properties of the \hi{} ring.

There is no diffuse stellar emission along the \hi{} ring in the CFHT Megacam image going down to a depth of $\sim$ 28 mag/arcsec$^2$. In cases where there is in-situ star-formation in the ring in the form of knots we expect to find objects with relatively blue colours embedded particularly on the high column density regions of the ring. They are also expected to have a diffuse and irregular stellar distribution resembling dwarf irregular galaxies. We hence visually inspected the region in and around the ring to search for the possible optical counterparts (OCs) to the ring. There are various objects coincident on the ring which mostly appear like background galaxies (which have evolved structures like bulges and discs). There are also several red barely resolved objects which are likely to be high-redshift galaxies. Interestingly, we also find several rather blue compact  OCs  close to  several high column density \hi{} condensations of the ring as denoted in Fig. \ref{HI_overlaid}. In Fig. \ref{OC_montage} we show a zoom-in of the five OCs. We obtained their magnitudes using aperture photometry in \textsc{iraf}. The magnitudes were then measured using circular apertures for all the OCs with a size of $\sim 3$ times the FWHM using the \textsc{phot} task in the \textsc{iraf} \textsc{appphot}. The background was determined by making an annulus around the OCs. Table \ref{tab:fluxes} summarises the positions and magnitudes of all the OCs of the ring. Note that in the i-band for all the OCs except for OC4 and in the r-band for OC2 and OC3 we could not get a reliable measure of the magnitude due to poor SNR. The $g - r$ colours for the OCs range from about 0.1 to 0.6 mag and are thus consistent with star-formation.

\begin{figure}
\begin{center}
\includegraphics[scale=0.55]{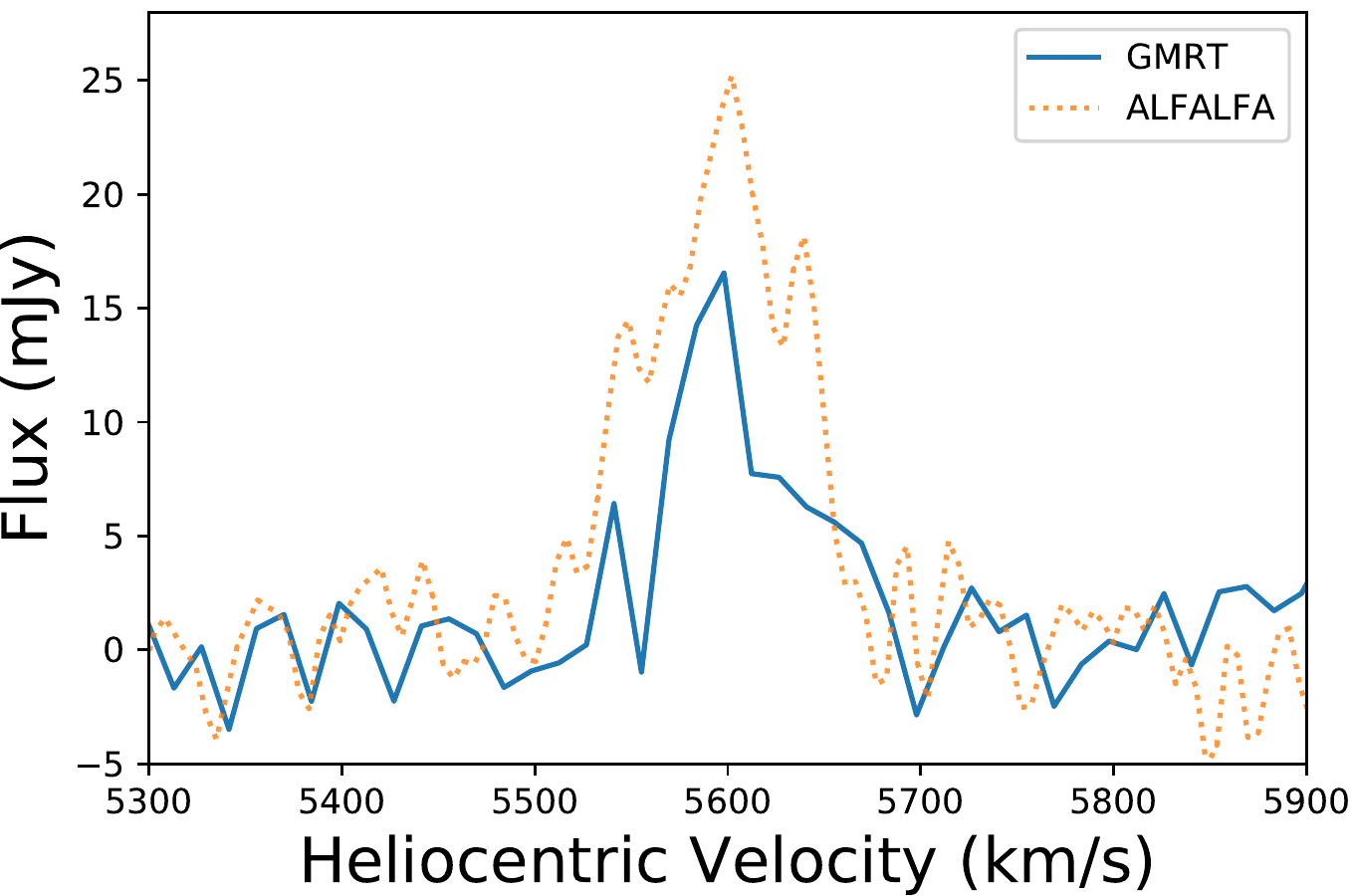} 
\caption{ Comparison of the integrated spectrum of the ring from GMRT (solid line) with the ALFALFA spectrum (dotted line). The GMRT spectrum has about 47\% of the flux missing which is quite uniformly spread across the velocity. See text for details.}  \label{spec_compare}

\end{center}

\end{figure}

 We estimated the integrated spectrum of the ring in our GMRT observations using the task \textsc{blsum} in \aips{}. The total detected \hi{} flux of the ring is 1.13$\pm$ 0.13 Jy~km/s, including the absolute 10\% calibrations errors. However, the total \hi{} flux from the single dish ALFALFA measurement around AGC 203001 is 2.14 $\pm$ 0.07 Jy~km/s. We thus miss out a significant fraction (47\%) of the flux in our GMRT observations. In Fig. \ref{spec_compare}, we compare the integrated spectrum of the ring using our GMRT observations (solid line) with that of the single dish ALFALFA measurements (dotted line). Notice that the missing flux is fairly uniformly distributed across velocity. Such a missing flux could be due to (1) calibration error (2) excessive flagging of short baselines or (3) a large fraction of the \hi{} is distributed in an extended diffuse component which is resolved out. The point source fluxes in our GMRT continuum image are in good agreement with the The NRAO VLA Sky Survey \citep[NVSS;][]{Condon98} fluxes. Moreover, there was minimal flagging done in the short baselines. A similar flux recovery was observed in a GMRT follow-up of the ALFALFA observations of the gas-rich interacting galaxy groups NGC 3166/9 and NGC 871/NGC 876/NGC 877 by \citet{Lee-Waddell12, Lee-Waddell14}. Hence we believe that there is still a significant amount of diffuse \hi{} surrounding AGC 203001 which is resolved out in our GMRT observations. A follow-up of this galaxy with a compact array (like the JVLA D-array) might help in detecting some of this flux.

\begin{figure*}
\begin{center}
\includegraphics[scale=0.62]{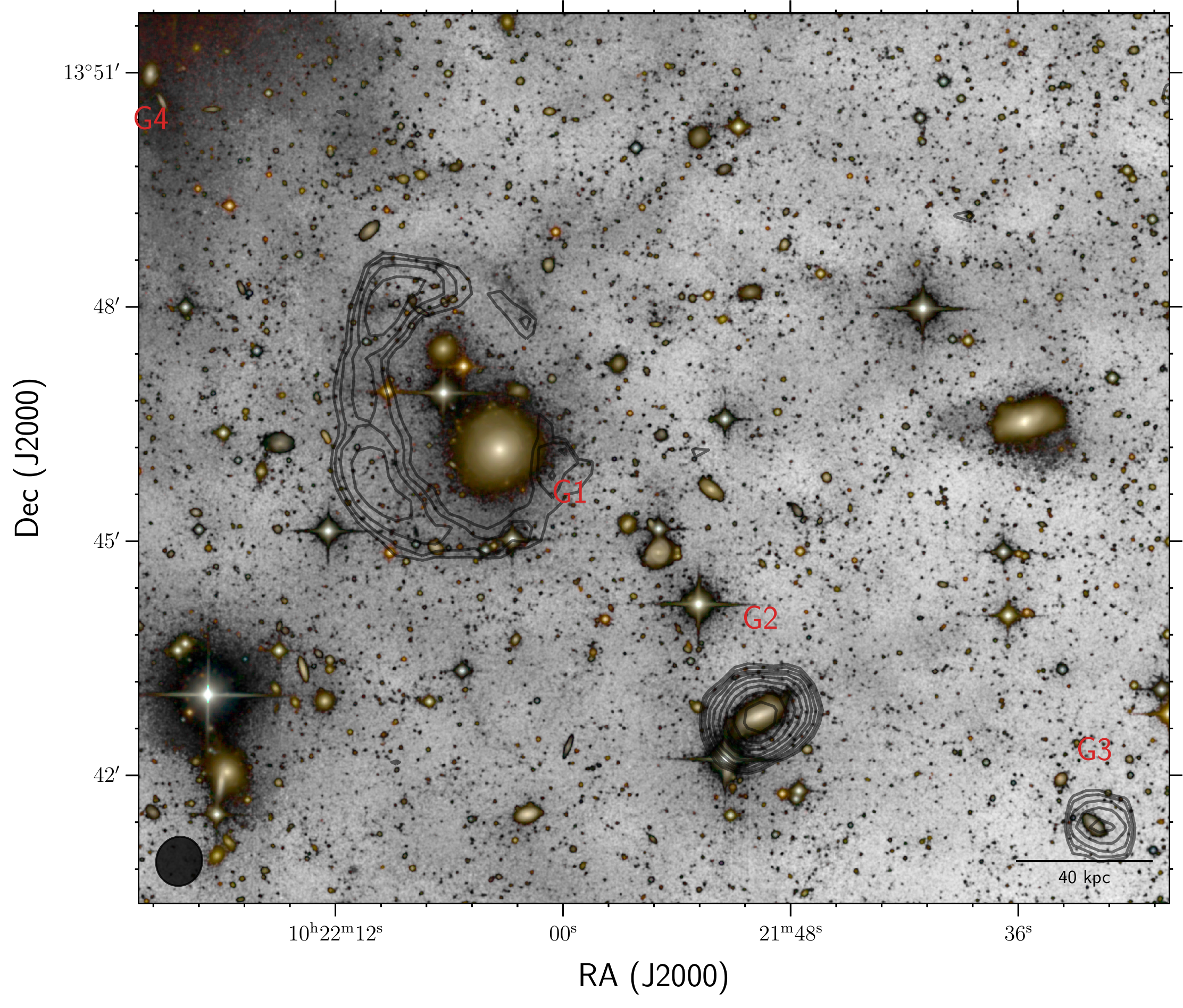}
\caption{Large scale image of the \hi{} ring overlaid on the CFHT  g-band image shown in reversed greyscale wherein the background is shown in white and emission is shown in grey. The $gri$ colour composite image is superimposed to identify the objects. The \hi{} contours are shown in black with the contour levels repeated from Fig. \ref{HI_overlaid}. Here we show the two more sources for which we detected \hi{}, denoted as G2 and G3. See text for more details.} 
\label{HI_largescale}
\end{center}
\end{figure*}

Fig. \ref{HI_largescale} shows the larger field of view of the \hi{} image of the ring and the galaxies around it. Here galaxies denoted by G2 (RA: 10h21m49.584s, DEC: +13d42m45.83s; AGC 205059), G4 (RA: 10h22m21.728s, DEC: +13d50m58.67s) and G1 (AGC 203001) are classified to be part of the same group based on the friends-of-friend group-finding technique \citep{Tempel14}. AGC 203001 along with G2 and G3 are also likely to be part of the loose group of three galaxies, WBL 263 \citep{White99} and USGC U292 with a virial radius of $\sim 0.72$~Mpc \citep{Ramella02}. \hi{} is detected in G2, a known  member of the group. Additionally, we also detect G3 (RA: 10h21m31.94s, DEC: +13d41m22.4s) in \hi{}. G3 is a dwarf galaxy which does not have spectroscopic data in the SDSS DR 15 \citep{SDSS_DR15}. From our \hi{} detection we obtain a systemic velocity of $\sim 5525$ km/s. Hence, G3 is at a similar distance as AGC203001 and likely part of the group. Notice that along the north-west section of the ring we see a curved LSB feature in the g-band image which appears to be associated with the ring. However, it is slightly misaligned with the ring and is possibly Galactic cirrus emission.

\section{Discussion} \label{discussion}
 In this section, we show how the \hi{} ring in this study compares with other known \hi{} bearing ETGs, in particular the \ATLAS{} galaxies \citep{Serra12}. We then briefly mention the possible formation scenarios for such a \hi{} dominated ring.
\subsection{Comparison with \hi{} observations of early-type galaxies}
\begin{figure}
\begin{center}
\includegraphics[scale=0.5]{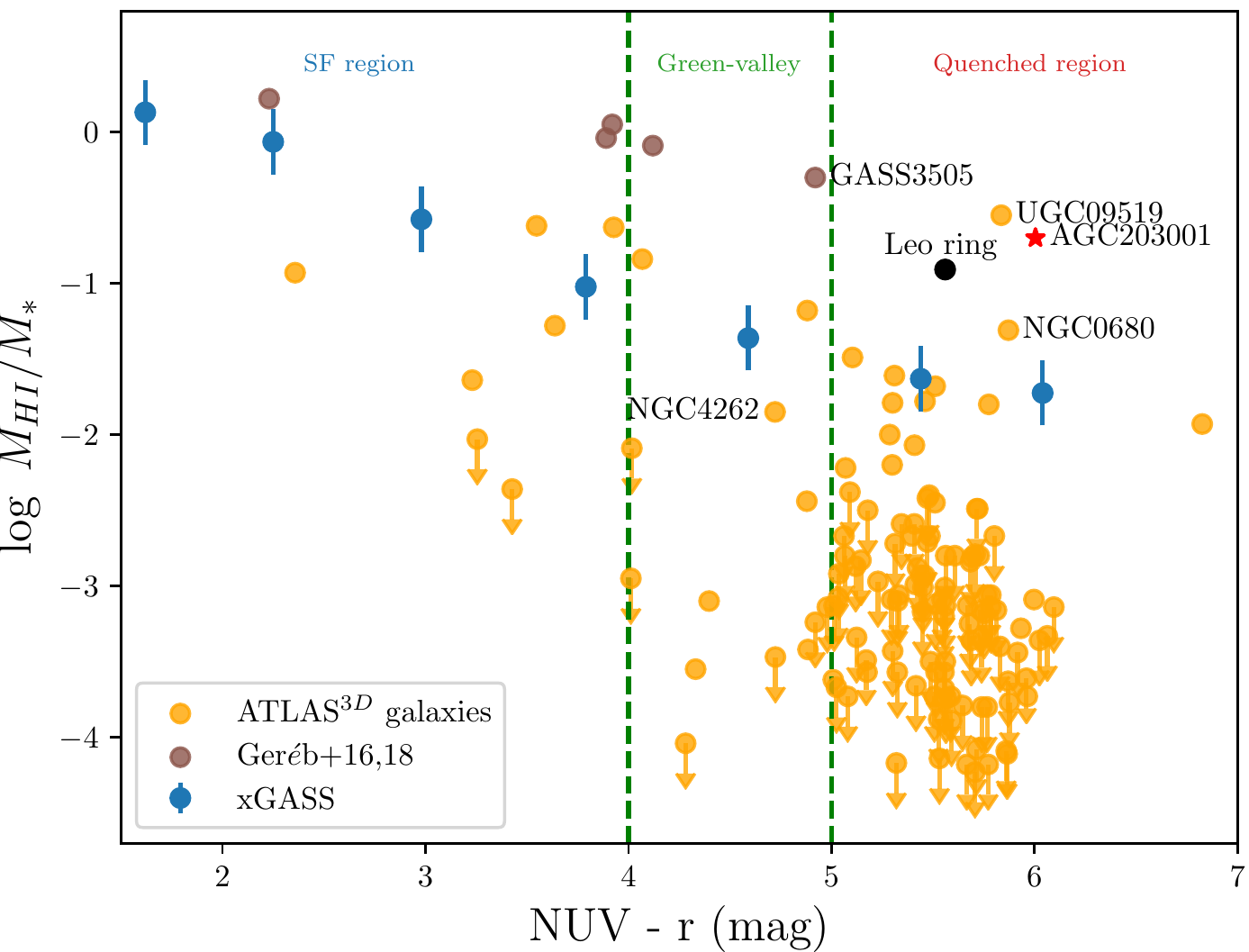}
\caption{ The \hi{} gas fraction vs. $NUV-r$ plane. The blue points represent the median values from the xGASS survey \citep{Catinella18}. AGC 203001 is marked with a red star and the Leo ring is marked with a black point. The yellow points represent the \ATLAS{} galaxies and the brown points represent the \hi{}-excess galaxies (with GASS3505 highlighted) from \citep{Gereb16, Gereb18}. The names of some of the \hi{}-rich \ATLAS{} galaxies, UGC09519 and NGC0680, and a ring galaxy NGC4262 are highlighted. The region within the green dashed line correspond to the green-valley from \citet{Salim2014}, and the region on the left and right corresponds to the SF and quenched region respectively. See text for discussion.} 
\label{HI_comparison}
\end{center}
\end{figure}
 Here we use the \hi{} gas fraction ($\log$(M$_{H{\sc i}}$/M$_*$)) vs. NUV-r plane \citep{Catinella18} to see how AGC 203001 compares particularly with \ATLAS{} galaxies (which forms a representative sample for ETGs with \hi{}) and other known \hi{} bearing ETGs from the literature. The details of data selection are provided in Appendix \ref{appendix:data}.

In Fig. \ref{HI_comparison}, we compare AGC 203001 (red star) with \ATLAS{} galaxies shown in yellow points on the \hi{} gas fraction vs. $NUV-r$~plane. For reference we show the median values from the mass-selected xGASS survey \citep{Catinella18}. The green dashed lines differentiate the star-forming (SF), green-valley and quenched region \citep{Salim2014}. Most of the \ATLAS{} galaxies, are gas poor, particularly the those in the quenched region. On the contrary, AGC 203001 is much more gas rich than \ATLAS{} galaxies and is in the quenched region (by virtue of its selection). And it is rather more close to the Leo ring, UGC 09519 and NGC 0680. We also show a sample of \hi{}-excess galaxies with very low star-formation efficiency from  \citet{Gereb16, Gereb18} (shown in brown), with GASS3505 highlighted. These galaxies were found to be outliers on the gas fraction vs. NUV -r colour and stellar mass surface density. Compared to AGC 203001, these galaxies are much more gas rich and are relatively more star-forming. We also highlight NGC 4262 (which is also part o the \ATLAS{} sample), a typical example of a barred lenticular galaxy with a \hi{} ring \citep{Krumm85}, which is classified as a polar ring galaxy by \citet{Khoperskov14}. In this case the \hi{} ring shows a much more vigorous star formation in the form of multiple star-forming knots \citep{Bettoni10}.

One of the galaxies, NGC 0680 which is as \hi{} rich as AGC 203001 is an interacting galaxy with an unsettled disc and thus has a very different morphology compared to AGC 203001 \citep{Serra12}. The other galaxy, UGC 09519, is classified to have a large \hi{} disc/ring morphology \citep{Serra12}. However, \hi{} disc/ring is well within the optical extent of the galaxy which is also very different from AGC 203001. Moreover, in the relatively deeper Mayall $z$-band Legacy Survey/Beijing-Arizona Sky Survey (MzLS/BASS) images \citep{Dey19} it clearly shows a diffuse optical ring unlike in AGC 203001. The other case is that of the radio galaxy B2 0648+27. It has very large amount of \hi{} mass ($\sim1.1\times 10^{10}$M$_\odot$) distributed in the form of a large ring/disc extending $\sim 160$ kpc and centered on the host galaxy \citep{Morganti03a}, as against AGC 203001. Moreover, the host galaxy shows clear signs of a major merger in the deeper optical image due to a distorted optical morphology and tidal arms \citep{Emonts08}. This again is different from the case of AGC 203001 presented here. We could not compare this galaxy on Fig. \ref{HI_comparison} due to lack of GALEX UV data. The Leo ring which is also an outlier in this gas scaling relation (close to AGC 203001) and also have a quenched host galaxy (even if we assume the host to be NGC 3379). More importantly, its \hi{} morphology  is in the form of a large off-centered ring extending $\sim 200$ kpc \citep{Schneider83, Schneider85}. It has no diffuse large-scale optical counterpart even in deep optical imaging \citep{Michel-Dansac10, Watkins14}. Only faint star-forming knots and diffuse structures were found within the ring in the UV \citep{Thilker2007} and in the optical \citep{Stierwalt09, Michel-Dansac10, Mihos18} suggesting in-situ star-formation. None of the galaxies in the Leo {\sc i} group show any signs of recent interactions, except for NGC 3384 which might have undergone a minor merger \citep{Watkins14}. 

Thus the \hi{} ring around AGC 203001 is unlike a typical \ATLAS{} galaxy, and has much more gas (due to its selection). In terms of its \hi{} morphology it is rather closer to other known cases of giant \hi{} rings around ETGs (e.g., the Leo ring and B2 0648+27). However, based on its optical properties it appears to most closely resemble the rare case of the Leo ring.


\subsection{Possible formation scenarios for the ring around AGC 203001}
The off-center nucleus of the \hi{} ring makes it a P-type ring based on the classification by \citet{Few86}. These types of rings are thought to have a collisional origin with an intruder galaxy which is typically within two diameters of the ring. We identify a nearby galaxy (G2) at a projected distance of $\sim 109$ kpc, which could be the possible intruder. A recent collision between G2 and the host galaxy at a relatively high speed could have triggered the formation of this ring. We also see asymmetrical outer isophotes in the optical image of the host galaxy which could be the signs of such an interaction. We note, however, that both the stellar and \hi{} components of G2 do not show any signs of recent interaction.


In some aspects, our ring is strikingly different from typical collisional ring galaxies. It lacks any bright optical counterpart suggesting almost negligible amounts of past and current events of SF. The high SFRs in the more traditional rings are due to gas compression after the collision. In a recent simulation of Cartwheel-like ring galaxies it was shown that SF first takes place in the expanding ring, soon after the collision, due to the induced density wave \citep[][]{Renaud18}. However, the \hi{} gas column density in our ring is much below the SF threshold (of $\sim$10$^{21}$ cm$^{-2}$), thus suggesting that the gas may not have been sufficiently compressed after the collision to induce  star formation. Very locally, the gas density was high enough though to allow the onset of limited star-formation, indicated by the presence of the blue OCs.


We propose a possible explanation and alternative scenarios for our observations as follows. It is possible that the host galaxy was a \hi{}-rich ETG, like those of the \ATLAS{} galaxies \citep{Serra12}, which  have low column density \hi{} gas disc to begin with and hence low SF, which underwent a collision to form the ring. This can explain the low SFR in the host galaxy and the ring. 
Another natural solution for all the discrepancies is that the possible collision of the gas between the intruder and AGC 203001, may have imparted a strong shock which can heat the gas in the host to very high temperatures a possibility pointed out by \citet{Appleton96}. Such a heating of the gas may have prevented SF in the ring, leading to a more diffuse \hi{} ring. The inclusion of such a shock heating in collisional ring galaxy simulations, an effect largely ignored until now, may provide extra parameter space to produce such a large and optically devoid \hi{} ring. 

Alternatively, as suggested by simulations in \citet{Bekki05}, such a ring-like structure could have formed due to tidal stripping of the outer-gas from a LSB galaxy having an extended \hi{} disc by the group potential. In their simulations, the \hi{} gas column density is expected to be quite low and it lacks any SF which is consistent with our observations of the \hi{} ring.

 Major merger is yet another possibility for the formation of such a ring around AGC 203001. \citet{Morganti03b, Morganti03a} proposed a evolutionary sequence wherein large \hi{}-rings/disc around ETGs could represent late stages of gas-rich mergers. In this scenario, the large-scale tidal tails around the Antennae galaxies (which is an ongoing merger) will eventually fall back in the centre forming a large diffuse disc/ring-like structure (like NGC 5266; \citep{Morganti97}). The host galaxy will undergo a starburst phase, followed by a possibly brief period of AGN activity (like B2 0648+27) and then eventually turn into a quenched ETG. In such a scenario, the ring around AGC 203001 would be placed in the late-stage of the evolutionary sequence. 

\section{Conclusion and Future Work}

 In conclusion, we have discovered an extremely rare example of a  large off-centered \hi{} ring, extending almost $\sim 115$~kpc in diameter around AGC 203001. Our deep CFHT $g$, $r$, and $i$-band images show that this ring has several faint optical counterpart at surface brightness levels of $\sim$ 28 mag/arcsec$^2$. Conventionally, ring galaxies are thought to have a collisional origin, although they also predict large amounts of SF in it, which is contrary to our observations. The origin of such \hi{}-dominated rings is still poorly understood.  In the future, we hope to increase the number of such extended \hi{} structures by mapping more galaxies found using our criteria which can help in understanding their formation scenario.

\section*{Acknowledgments}
We thank the anonymous referee for several insightful comments that have improved both the content and presentation of this paper. OB would like to thank Nissim Kanekar for help with the GMRT data analysis. We thank the staff of the GMRT who have made these observations possible. GMRT is run by the National Centre for Radio Astrophysics of the Tata Institute of Fundamental Research. Based on observations obtained with MegaPrime/MegaCam, a joint project of CFHT and CEA/DAPNIA, at the Canada-France-Hawaii Telescope (CFHT) which is operated by the National Research Council (NRC) of Canada, the Institut National des Sciences de l'Univers of the Centre National de la Recherche Scientifique of France, and the University of Hawaii. PK's work at RUB is partly supported by BMBF project 05A17PC2 for D-MeerKAT. This research made use of APLpy, an open-source plotting package for Python \citep{aplpy:12}. This research made use to matplotlib \citep{Hunter:2007}. This research made use of Astropy,\footnote{http://www.astropy.org} a community-developed core Python package for Astronomy \citep{astropy:2013, astropy:2018}. This research has made use of NASA's Astrophysics Data System.

\bibliographystyle{mnras}
\bibliography{HI_ring.bib}

\include{appendix}
\end{document}

%% file: appendix.tex
\appendix
\section{Details of the data selection for the comparison with other \hi{} bearing ETGs} \label{appendix:data}

For AGC 203001 and \ATLAS{} galaxies the $NUV$ magnitudes are from the revised catalogue of GALEX UV sources from \citet{Bianchi17}. For NGC 0770, NGC 3458 and NGC 4262 the NUV magnitudes were missing from the \citet{Bianchi17} catalogue and were hence taken from the older \cite{GildePaz07} catalogue. The NUV magnitudes are corrected for Galactic extinction following \citet{Wyder07} with E(B - V) values from the \citet{Schlegel98} dust maps. For \ATLAS{} galaxies the extinction corrected SDSS $r$-band magnitudes are taken the catalogue of nearby ETGs \citep{Dabringhausen16}. The stellar masses for the \ATLAS{} galaxies are also taken from \citet{Dabringhausen16}. The stellar mass and $r$-band magnitude for AGC 203001 are taken from GSWLC \citep{Salim16} and SDSS DR15 \citep{Alam2015} respectively. The \hi{} masses for the \ATLAS{} galaxies are from \citet{Serra12}. We use the \hi{} mass for AGC 203001 from the single dish ALFALFA survey since in our GMRT observations we did not recover all the flux. \citet{Michel-Dansac10} show that the Leo ring most likely formed due to a head-on collision between NGC 3384 and M 96. Hence for the Leo ring, we assume that NGC 3384 is the host galaxy and use its stellar mass and $r$-band magnitude from \citet{Dabringhausen16} and $NUV$ magnitude from \citet{Bianchi17}. Assuming a distance of $11.574$ Mpc (same as that of NGC 3384) for the Leo ring, we take its \hi{} mass to be 2.14 $\times$ 10$^9$M$_\odot$ \citep{Schneider83}.